# Influence of the vicinal substrate miscut on the anisotropic two-dimensional electronic transport in $Al_2O_3$-$SrTiO_3$ heterostructures

K. Wolff, R. Schäfer, D. Arnold, R. Schneider, M. Le Tacon, and D. Fuchs[a]

Karlsruhe Institute of Technology, Institute for Quantum Materials and Technologies,
D-76021 Karlsruhe, Germany

The electrical resistance of the two-dimensional electron system (2DES) which forms at the interface of $SrTiO_3$ (STO)-based heterostructures displays anisotropic transport with respect to the direction of current flow at low temperature. We have investigated the influence of terraces at the surface of STO substrates from which the 2DES are prepared. Such terraces are always present in commercially available STO substrates due to the tolerance of surface preparation which result in small miscut angles of the order of $\gamma \approx 0.1°$ with respect to the surface normal. By a controlled increase of the substrate miscut we could systematically reduce the width of the terraces and thereby increase the density of substrate surface steps. The in-plane anisotropy of the electrical resistance was studied as a function of the miscut angle $\gamma$ and found to be mainly related to interfacial scattering arising from the substrate surface steps. However, the influence of $\gamma$ was notably reduced by the occurrence of step-bunching and lattice-dislocations in the STO substrate material.

Magnetoresistance (MR) depends on the current orientation as well, reflecting the anisotropy of carrier mobility. For $\gamma \geq 2°$, MR is substantially enhanced and shows the trend towards a linear field dependence which is typical for inhomogeneous conductors. From weak-antilocalization observed at small magnetic field we deduce information on inelastic scattering and spin-orbit coupling. While the field scale associated with a Rashba-type spin-orbit coupling in 2D weak-localization does not show a pronounced correlation with $\gamma$, distinct changes of the scale are associated with inelastic scattering.

[a] Author to whom correspondence should be addressed: dirk.fuchs@kit.edu

*Data Availability Statement*: The data that supports the findings of this study are available within the article.

## I. Introduction

The discovery of interfacial two-dimensional (2D) superconductivity in oxide heterostructures, whose building blocks consist of band-insulators, has attracted a lot of attention [1,2]. One of the most studied systems is arguably the conducting interface between $LaAlO_3$ (LAO) and $SrTiO_3$ (STO) perovskites [2], sketched in Fig. 1a, most notably due to the emergence of *e.g.* superconductivity [3], multiple quantum criticality [4], magnetism [5], or tunable Rashba-type spin-orbit coupling (SOC) [6]. As such, LAO/STO has rapidly become a canonical system for studying the impact of electronic correlations in 2D [7-10].

The physics of the confined *d*-orbital states is generally more complex, but also far richer, than that encountered in the case of *p*-states of electron gases in semiconductors. Tunneling experiments on LAO/STO for instance indicate that this system behaves like a correlated 2D electron liquid (2DEL), rather like a 2D electron gas (2DEG) [11]. This is for instance illustrated by the observation of a Berezinskii–Kosterlitz–Thouless (BKT) transition (at $T_{BKT}$), indicative for a 2D superconducting state, just below the onset of superconductivity (at $T_c$). Nevertheless, it has been suggested that the resistive tail of LAO/STO below $T_{BKT}$ might be related to the presence of inhomogeneities rather than to finite-size effects [12]. For example, the superconducting transition in LAO/STO can be phenomenologically well described by the percolation of filamentary structures of superconducting "puddles" with randomly distributed transition temperatures, embedded in a non-superconducting matrix [13]. Mesoscopic inhomogeneities were also found by direct measurements of the superfluid density [14] and magnetism [15,16]. Anisotropic transport properties have been reported and assigned to extrinsic effects such as step-edges caused by the terraced surface structure of the $TiO_2$-terminated STO substrates [17,18], a net surface charge at the step edges [19] or structural domains of STO [20,21]. However, the observation of a negative compressibility of the 2DEL [15] also hints at an intrinsic mechanism that results in charge segregation and electronic phase separation even in a perfectly clean and homogeneous system. Furthermore, in strong magnetic fields, anisotropic transport has been suggested to arise naturally alike, as a consequence of the orbital character of the sub-bands [22,23,6].

Anisotropic electronic transport properties are also observed in other 2D interfacial electron systems such as $LaTiO_3$/STO [4] or $Al_2O_3$/STO (AO/STO) heterostructures [24] and are obviously not unique to the 2DEL in LAO/STO. The resistance anisotropy of AO/STO heterostructures [24] might be a fingerprint for a filamentary structure and strongly suggests that interfacial superconductivity is well defined within the individual terraces which are only weakly coupled to each other - highlighting the importance of the surface structure for the anisotropic in-plane electronic transport. Therefore, surface properties of the used substrate materials may be critical not only with respect to lattice mismatch and resulting film strain in case of epitaxial growth but also with respect to the substrate miscut angle and resulting stepped-terraces on the substrate surface which may induce discontinuities or distinct line-defects in the film plane. This offers new perspectives for the control of the anisotropies of interfacial electronic transport in STO-based heterostructures, and calls for a detailed exploration of the impact of stepped terrace surfaces and of the step-edges of vicinal substrates. Furthermore, this opens interesting new perspective, such as the possibility of a "giant" Rashba effect. In LAO/STO, the polar field and symmetry breaking at the interface results in a Rashba-type spin orbit coupling (SOC) whose strength is related to the number of charge carriers and electric-field strength at the interface. The Rashba effect was recently brought to attention because of the possibility of spintronics without a magnetic field [25,26]. Usually, the spin-splitting in most 2D electron systems is not sufficient for applications. However, lower dimensional electron systems may display strong SOC and merit investigations into spin manipulation [27,28].

One can typically expect electronic anisotropy of AO/STO heterostructures to increase with decreasing terrace width *w*, which relates to the miscut angle $\gamma$ of the substrate through $w = a/\tan\gamma$, where $a$=3.905 Å is the unit-cell height of STO (see Fig. 1(b)). Here, we report on a systematic study on the influence of the vicinal substrate miscut of the $TiO_2$-terminated STO substrates on the anisotropic 2D electronic transport in AO/STO heterostructures. To this end, the vicinal miscut angle $\gamma$ which were used for the preparation of the heterostructures, was varied from 0.1° to 6°. The sheet resistance was measured as a function of temperature *T* and magnetic field *B* for different in-plane directions and analyzed with respect to the dominant scattering mechanisms and SOC. Anisotropic transport appears below about 30 K and is significantly enhanced for $\gamma > 0.5°$. However, a distinct relation of the Rashba-type SOC to the miscut angle $\gamma$ could not be verified, especially no giant Rashba effect.

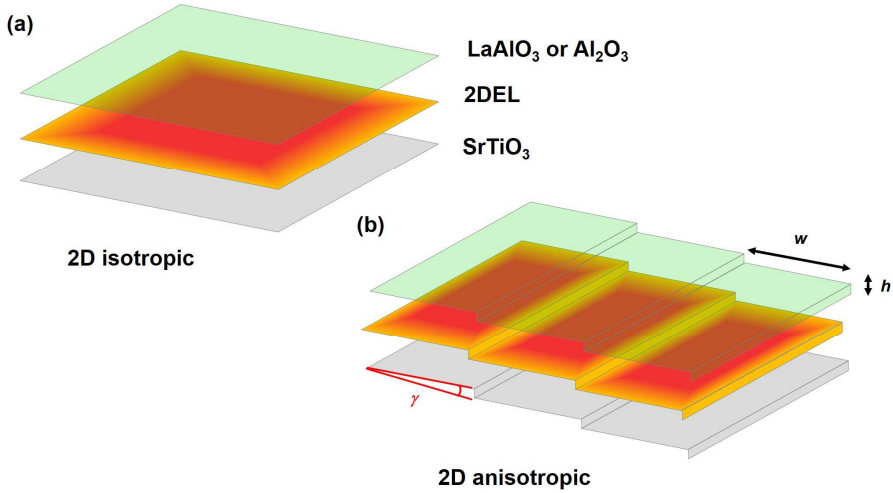

Figure 1: Schematic of a 2DEL at the interface of $TiO_2$-terminated STO-based heterostructures, e. g., LAO/STO or AO/STO. (a) Idealized atomically flat interface with no disruptions and isotropic surface structure. (b) The interface displays discontinuities, *i. e.*, terraces with a width $w$ and a step-height $h$ of one STO unit-cell, caused by the miscut of the substrate. The discontinuities may lead to a reduced conductivity at the step edges, indicated by the color gradient, and therefore responsible for anisotropic electrical transport. The substrate miscut angel $\gamma$ is given $\tan\gamma = h/w$.

## II. Experimental

The key idea of the work was to tune electrical anisotropy in STO-based 2DEL by using (001) oriented $TiO_2$-terminated substrates (Ti-STO) with different vicinal miscut angle $\gamma$. To this end, STO substrates with $\gamma$ = 0.1°, 0.5°, 2°, and 6° ($w \approx$ 224 nm, 45 nm, 11 nm, and 4 nm, respectively) from CrysTec GmbH were used.

The preparation of atomically flat terraces on the substrate surface is an indispensable step to prepare well-characterized samples [29]. Typically, a Ti-STO substrate surface is achieved by using acid based etchant [30-33] or deionized water treatment [34] and subsequent annealing which may vary with $\gamma$. Here, we used selective wet chemical etching in buffered HF acid (BHF) of the substrate surface to obtain $TiO_2$-termination of the substrate surface. To recrystallize the etched surface, the substrates were annealed under flowing oxygen. This annealing step was essential to obtain well defined terraces at the surface and had to be optimized with respect to annealing temperature and -time for the various miscut angles, *i. e.*, terrace widths. The surface topography was characterized by atomic force microscopy (AFM). In Fig. 2 we have shown the surface topography of an untreated (a), only annealed (b), and BHF-etched and annealed (c) STO substrate. Well defined terraces with a step height of one STO unit-cell have been obtained after optimization of selective etching time and annealing temperature $T_{an}$ and time $t_{an}$. The recrystallization is a thermally activated diffusion process. The mean particle displacement or surface diffusion length is - according to Fick`s law - proportional to $(D \times t)^{1/2}$, where $D$ is the diffusion constant and $t$ the time. According to the Arrhenius law, $D$ increases exponentially with $T$. That results in an increase of the annealing time $t_{an} \sim w^2/D$ with increasing terrace width and decreasing $D$ or $T$. $t_{an}$ too short usually leads to meandering of the terraces, while for $t_{an}$ too large unwanted step bunching may occur. We have varied $T_{an}$ from 800°C – 950°C and $t_{an}$ from 0 – 120h. The annealing temperature $T_{an}$ could not be increased above 950°C because of the segregation of Sr and the formation of precipitates at the surface. Since $t_{an}$ increases $\sim e^{-T}$, we avoided to lower $T_{an}$ below 800°C. Figures 2(d-g) demonstrate the Ti-STO surface topography of the vicinal substrates after optimization. The terraces are aligned perpendicular to the miscut direction and display same orientation throughout the substrate surface. Surprisingly, $T_{an}$ and $t_{an}$ were very similar for the various vicinal substrates, i. e., $T_{an} \approx 950°$ and $t_{an} = 5 - 10$ h. For the high miscut substrates, $\gamma = 2°$ and $6°$, some step-bunching occurred which could not be avoided by the variation of $T_{an}$ and $t_{an}$, see Figs. 2(f, g). The step-bunching resulted in a mean step-height larger than 3.905 Å and hence a terrace width $w_{exp}$, which is the average value over about 20 terraces, larger than the expected width $w_{theo}$, as calculated from the miscut angle $\gamma$, see Fig. 3.

To characterize electrical transport along specific crystallographic directions ($\varphi$), especially with respect to the terraces/step edges, we produced microbridges (20×100 μm$^2$) and Hall-bars by standard ultra-violet photolithography and a CeO$_2$ hard-mask technique [35,18], see Figs. 2(h, i). The films were deposited by pulsed laser deposition of Al$_2$O$_3$ at a substrate temperature of 250°C in an oxygen atmosphere with partial pressure $p(O_2) = 10^{-6}$ mbar, more details are described elsewhere [35,18]. The crystal structure was found to be strongly related to that of the cubic defect spinel-type $\gamma$-Al$_2$O$_3$ [36]. For electrical transport measurements, the microbridges were contacted by ultrasonic Al-wire bonding. Measurements of the sheet-, Hall-, and magneto-resistance were carried out in a physical property measurement system (PPMS) from Quantum Design.

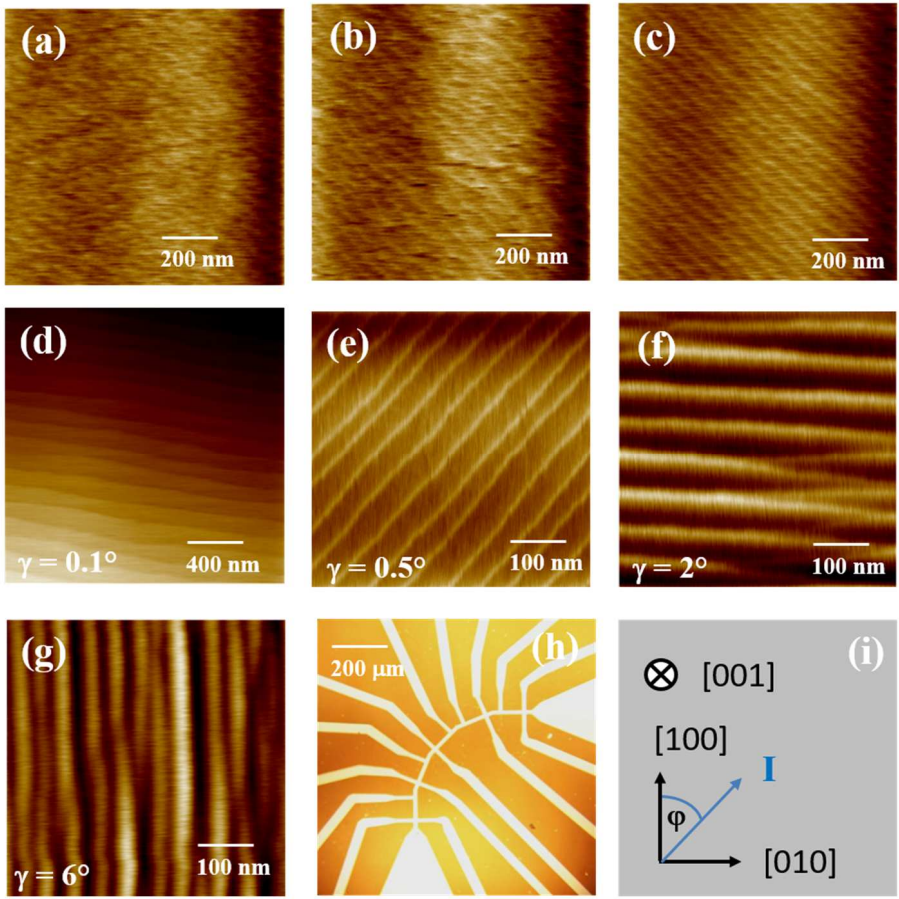

Figure 2: Surface topography of STO substrates characterized by AFM. (a) STO as delivered from CrysTec company. (b) STO after 5h annealing at 950°C. (c) STO after BHF-etching and annealing. (d-g) Surface topography of the vicinal STO substrates ($\gamma$ = 0.1°, 0.5°, 2°, and 6°) after optimization of $T_{an}$ and $t_{an}$. (h) Microbridges (20×100 μm), bright areas, produced by standard photolithography and a CeO$_2$ hard mask technique. (i) Crystallographic orientation of the shown substrates and definition of $\varphi$, the angle between [100]- and current flow direction.

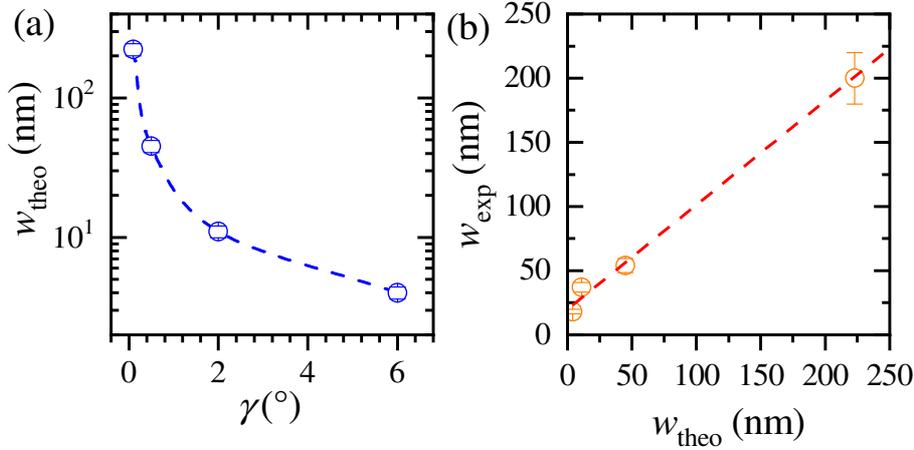

Figure 3: (a) Terrace width $w_{theo}$ as calculated from the miscut angle $\gamma$ assuming unit-cell step-height of the terraces versus $\gamma$. (b) Terrace width $w_{exp}$, the average value over about 20 terraces, as determined by AFM measurements versus $w_{theo}$, as calculated from the miscut angle $\gamma$. The intersection at $w_{exp} > 0$ is caused by step bunching.

III. Results and Discussion

A. Anisotropic transport of the 2DEL in AO/STO heterostructures

Figure 4 gives an overview of the sheet resistance $R_s$ versus $T$ behavior of the various heterostructures grown on vicinal STO substrates. For each sample, $R_s$ vs. $T$ is shown for different current directions $\varphi$. Generally, above $T = 100$ K, $R_s$ does not depend on $\varphi$ and shows isotropic, almost $T^2$-dependence, which is attributed to strongly renormalized electron-phonon interaction [37,38]. Below a shallow minimum around 30 K, $R_s$ increases again while becoming clearly anisotropic. For 2 K ≤ $T$ ≤ 5 K, $R_s$ stays nearly constant. The resistance ratio of $R_s(300K)/R_s(5K)$ amounts to about 20 for all the heterostructures. The overall $R_s(T)$ behavior can be well explained by combining electron-phonon interaction at high temperatures and impurity scattering at low temperatures. For impurity scattering, charge carrier trapping and electrostatic screening by the large dielectric permittivity of STO were taken into account [38]. Fits of $R_s$ with respect to that model are shown in Fig. 4 by solid lines.

Electric transport anisotropies develop below 30 K, where impurity scattering dominates transport. Beside statistically distributed scattering centers which lead to an isotropic contribution, $R_{iso}$, the residual resistance at 5K also comprises anisotropic contributions. Mainly responsible for that are very likely dislocation lines in the bulk of STO and defect scattering at step-edges. It is well known, that the flame fusion growth technique of STO single crystals results in a high density of <110> lattice-dislocations in the bulk material [39]. A preferential orientation of dislocations lines along <110> lattice-direction results in an increased scattering rate and hence resistance contribution $R_d$ perpendicular to those defect lines. Of course, defect scattering at the step-edges, *i. e.*, interfacial scattering, must be considered [17]. Although only restricted to the interface, it might become significant in 2DEL with a thickness of only a few nanometers. For well-defined and aligned step-edges, the scattering rate is enhanced perpendicular to the step-edges/terraces, resulting in an anisotropic contribution $R_t$ to the residual resistance. $R_t$ is expected to increase with the density of step-edges, terraces, or the miscut angle $\gamma$.

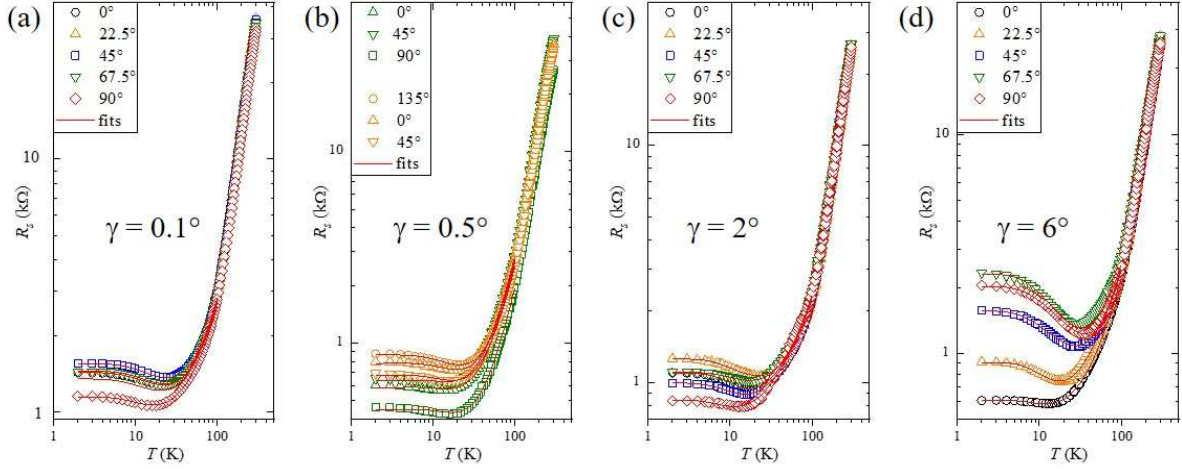

Figure 4: In-plane sheet resistance $R_s$ for various current-flow directions versus temperature $T$ for AO/STO heterostructures grown on vicinal TiO$_2$-terminated STO substrates with miscut angle $\gamma = 0.1°$ (a), 0.5° (b), 2° (c), and 6° (d). The current-flow direction is given by the angle $\varphi$ with respect to the [100] direction of STO, see legend. Solid lines indicate fits to the data assuming electron-phonon and impurity scattering, for details see [18, 38].

In the following, we have modeled the $\varphi$-dependence of the residual resistance by: $R_s(\varphi) = R_{iso} + R_d + R_t$ with $R_d = r_d \times \sin^2(\varphi-\omega_d)$ and $R_t = r_t \times \sin^2(\varphi-\omega_t)$, where $r_d$ and $r_t$ are the amplitudes of $R_d$ and $R_t$, and $\omega_d$ and $\omega_t$ the angles of the dislocation lines and terraces, respectively, with respect to the [100]-direction of STO. In-plane anisotropy for step-edges and terraces are assumed to be two-fold and homogeneous distributed. Hence $\omega_i = \omega_i+180°$. However, if $\omega_t = \omega_d+90°$ or $\omega_d-90°$ it is not straightforward to deduce distinct contributions $R_t$ and $R_d$. In that case, additional measurements are necessary. To this purpose, customized substrates from CrysTec company were used with vicinal miscut direction along the [110] and [1-10] direction without changing crystal orientation. This was for example necessary for $\gamma = 0.5°$ (see Fig. 5(b)) where we have used substrates from the same batch with $\omega_t = \omega_d - 90°$ and $\omega_t = \omega_d +180°$ provided by CrysTec Company.

In Fig. 5 we have documented the $\varphi$-dependence of the total sheet resistance at $T = 5$ K as well as the isotropic contribution $R_{iso}$ and the anisotropic contributions $R_d$ and $R_t$. The polar plot documents the anisotropic behavior of $R_s$ which is strongest for the largest miscut angle $\gamma = 6°$. As expected, $R_t$ displays maxima for current flow direction perpendicular to the terraces and steadily increases with $\gamma$, i. e., with the number of step-edges/terraces per channel length. However, the anisotropy is not only caused by $R_t$ but also significantly affected by $R_d$ as well. For the samples shown in Fig. 5, $R_d$ displays maxima at $\varphi = 45°$ and 225° ($\omega_d = 135°$ or 315°) as expected for <110> dislocation lines. Depending on the substrate batch, maxima of $R_d$ at $\varphi = 135°$ and 315° ($\omega_d = 45°$ or 225°) have been observed for other samples alike. $R_d$ dominates anisotropy for the first sample where $\gamma = 0.1°$ is smallest, see Fig. 5(a). Obviously, the anisotropic distribution of dislocation lines varies for the different vicinal substrates. Note that simultaneous occurrence of dislocation lines at $\omega_d$ and $\omega_d +90°$, will cancel out anisotropic scattering and contribute to $R_{iso}$ only. This may explain the variation of $R_d$ and $R_{iso}$.

The main result of the analysis is the steady increase of $r_t$ with increasing miscut angle $\gamma$, see Fig. 6(a). The anisotropic contribution to the residual resistance by $R_t$ can be increased by about one order of magnitude when $\gamma$ is increased from 0.1° to 6°. As expected, the experiment clearly demonstrates the tunability of the anisotropic resistance $R_t$ by surface engineering. However, the control of $R_t$ by $\gamma$ is likely limited due to the occurrence of step-bunching after surface treatment for substrates displaying large miscut ($\gamma \geq 2°$), see Fig. 2(f,g). Step-bunching results in an increase of $w$ and therefore to a reduction of the expected density of step-edges, reducing slightly the influence of $\gamma$ on $R_t$. Furthermore, an anisotropic distribution of noncontrollable defects, such as <110> lattice dislocations, may also add significantly to the anisotropic transport and therefore may diminish the influence of $R_t$ on the overall resistance anisotropy. In Fig. 6(b) we have shown the maximum resistance anisotropy $R_{max}/R_{min}$, where $R_{max}$ and $R_{min}$ are the maximum and minimum $R_s(\varphi)$-values at $T = 5$ K, as a function of the miscut angle $\gamma$. Since $R_d(\varphi)$ also enhances or even dominates the anisotropy for some samples, $R_{max}/R_{min}$ does not necessarily increase with increasing $\gamma$, which prevents accurate control of resistance anisotropy by $\gamma$.

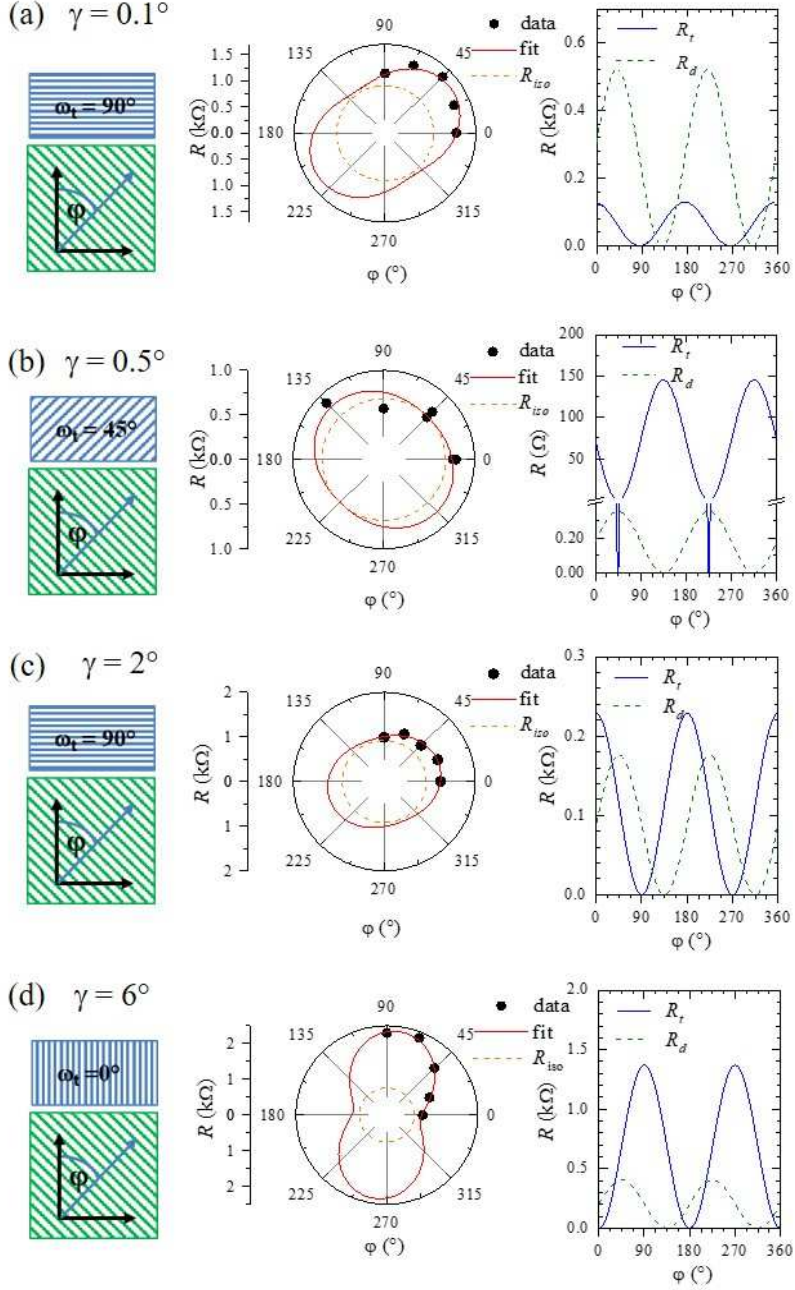

Figure 5: Residual resistance at $T = 5$ K as a function of the current direction $\varphi$ for AO/STO heterostructures on vicinal STO substrates with miscut angle $\gamma = 0.1°$ (a), 0.5° (b), 2° (c), and 6° (d). Data from (a) are the same as shown in Fig. 3(a) of Ref. [18]. Direction of the step edges/terraces, $\omega_t$, is indicated by the top blue shaded area on the left. The direction of the dislocation lines is indicated by the green bottom shaded area where also the direction of the current $I$ which is given by the angle $\varphi$, is indicated. The polar plot in the middle displays the $\varphi$-dependence of the total resistance $R_s$, the corresponding fit to the data, and the isotropic contribution $R_{iso}$ to $R_s$. The $\varphi$-dependence of the anisotropic contributions $R_t$ and $R_d$ is shown on the right. Data for $\gamma = 0.5°$ and $\omega_t = 315°$, which we used in addition to deduce $R_t$ and $R_d$, are not shown in (b). The amplitudes $r_t$ as deduced from the fits are 0.12, 0.15, 0.22, and 1.4 k$\Omega$ for $\gamma = 0.1°$, 0.5°, 2°, and 6°, respectively.

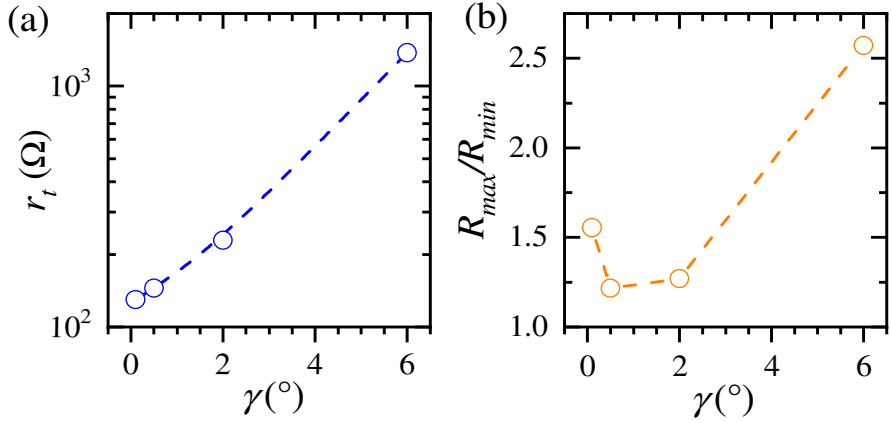

Figure 6: (a) Amplitude $r_t$ versus miscut angle $\gamma$. The anisotropic contribution to $R_s$ due to impurity scattering by interfacial step edges increases with increasing step density and hence miscut angle $\gamma$. (b) The maximum resistance anisotropy $R_{max}/R_{min}$ at $T = 5$ K as a function of $\gamma$. $r_t$, $R_{max}$, and $R_{min}$ were obtained from fits to $R_s(\varphi)$, see text.

B. Rashba-type spin-orbit coupling (SOC) in STO-based 2DEL

The Rashba effect is a manifestation of spin-orbit interaction in solids, where spin degeneracy associated with the spatial inversion symmetry is lifted due to a symmetry-breaking electric field normal to the heterointerface [40]. Anisotropic electronic transport may indicate a lowering of symmetry at the interface and hence result in a change of the Rashba-type SOC. Especially, surface steps may generate additional symmetry breaking and electric fields which influence Rashba-type SOC. Therefore, we have carried out a detailed investigation of the SOC in polar AO/STO heterostructures as a function of the electronic anisotropy alike. The electronic anisotropy was varied, as shown in the previous section, by the substrate miscut angle $\gamma$. The Rashba effect was studied by weak localization or antilocalization (WAL) analysis of the low-temperature magnetoresistance (MR). The analysis of WAL was carried out using the Maekawa-Fukuyama theory [41]. A detailed description of the analysis is given elsewhere [18].

In Fig.7 we show the magnetoresistance $MR = [R_s(B)-R_s(0)]/R_s(0)$ versus $B$ at $T = 10$ K for AO/STO heterostructures with $\gamma = 0.1°$, $0.5°$, $2°$, and $6°$ for the various current-flow directions $\varphi$. Generally, $MR(B,T)$ is always positive and increases with increasing $B$ and decreasing $T$ reaching maximum values of 5 - 25 %. Usually, quantum correction effects such as WAL are already absent to the most part for $T \geq 10$ K so that MR is dominated by classical Lorentz scattering (LS). Obviously, MR displays anisotropic behavior with respect to $\varphi$ as already observed for the residual resistance $R_s(0)$. According to Kohler`s rule for small magnetic fields, classical MR caused by LS increases proportional to the square of the charge carrier mobility $\mu$ [42]. The sheet carrier concentration $n_s \approx 2\times10^{13}$ cm$^{-2}$ as deduced from the linear Hall resistance was found to be well comparable for all the heterostructures so that $\mu$ is proportional to the inverse of $R_s(0)$ expecting $MR \sim 1/R_s(0)^2$. Consistently with the anisotropic behavior of $R_s(0)$, see Figs. 4 and 5, for a given miscut angle $\gamma$ the largest MR is always observed for current flow direction where $R_s(0)$ is smallest. The $\varphi$-dependent anisotropy of MR correlates fine with $R_s(0)$. Therefore, the measurements strongly suggest, that the anisotropic behavior of MR for $T \geq 10$ K is likewise dominated by defect scattering on step-edges and lattice dislocations.

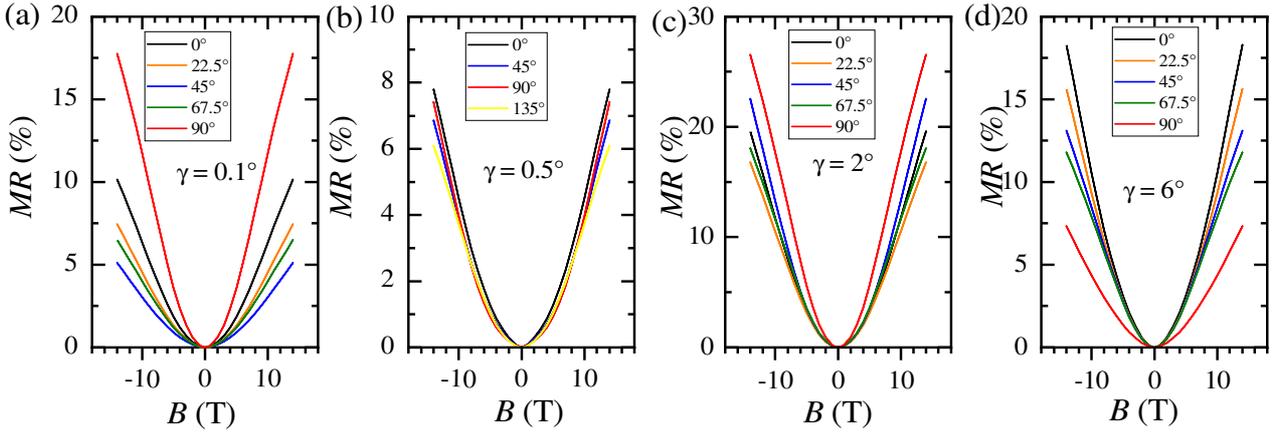

Figure 7: Magnetoresistance *MR* versus magnetic field *B* for various current-flow directions at *T* = 10 K of AO/STO heterostructures prepared on vicinal STO-substrates with miscut angle γ = 0.1° (a), 0.5° (b), 2° (c), and 6° (d). The current-flow direction is given by the angle φ with respect to the [100] direction of STO, see legend. *B* was applied parallel to the surface normal.

In Fig. 8 we have displayed *MR* versus *B* at *T* = 2 K for the AO/STO heterostructures with γ = 0.1°, 0.5°, 2°, and 6°. The measurements were carried out for current flow direction at φ = φ*, which is the direction where $R_s(φ)$ displays its maximum value $R_{max}$, cf. Fig. 5.

For *T* < 5 K, MR displays a dip at small magnetic fields, typically related to quantum corrections due to WAL. The maxima of *MR* are comparable for γ = 0.1° and 0.5° and distinct smaller compared to those of heterostructures grown on the high miscut substrates with γ = 2° and 6°. However, this is obviously not the case for all samples and may be related to the enhanced anisotropic conduction for γ ≥ 2°. Disorder-induced charge carrier density inhomogeneities in 2D electron systems may indeed result in an enhancement of *MR* [43]. It has been shown that *MR* scales quadratically with the ratio $n_{rms}/n_0$, where $n_0$ and $n_{rms}$ is the charge carrier density and fluctuation, respectively. Furthermore, *MR* likewise displays a quadratic field dependence. In the high magnetic field regime, μ*B* >> 1, a linear field-dependence has been deduced for 2D disordered systems [44]. For γ = 6°, *MR* indeed displays a more linear field dependence which might be explained in that context by the smallest μ ≈ 100 cm$^2$/(Vs) of all the heterostructures. Therefore, the linear *MR* is very likely a further indication for the inhomogeneous conductivity in that sample [44]. Because of the rather low charge carrier mobility of the samples, we do not expect quantum effects [45], mobility- [46] or density-fluctuations [47] of high-mobility electrons to be responsible for the linear *B*-dependence of *MR*. A linear band-dispersion or Dirac-like electrons which may give rise to a giant linear *MR* [10] can be very likely excluded as well for the AO/STO heterostructures here.

In the low *T* and *B* regime of *MR(B,T)*, a dip, *i. e.*, an additional positive contribution to the classical part of *MR* appears, indicative for WAL [48]. For STO-based 2DEL, WAL is most pronounced at low temperatures and small fields and vanishes for *T* > 10 K and *B* > 3 T due to the loss of electronic phase coherence [6,48]. The field dependence of *MR* is indeed well described by classical electron scattering and additional contributions from WAL [42,41,18], see fits (solid lines) to the data in Fig. 8. A detailed description of the fitting analysis is given elsewhere [18]. Obviously, relative contributions from WAL to *MR* are strongest for γ = 0.1° and 0.5° and seem to be less pronounced for γ = 2° and 6°. Since the absolute contribution to the conductivity by WAL is always of the order of $e^2/h$, where *e* is the elementary charge and *h* the Planck constant, the reduction of the relative contributions by WAL for γ = 2° and 6° may be simply related to the disorder-induced enhancement of *MR*. However, disorder may likewise lead to enhanced charge carrier localization and hence to the disappearance of WAL. A more detailed analysis of WAL especially with respect to the SOC in the heterostructures is given in the following.

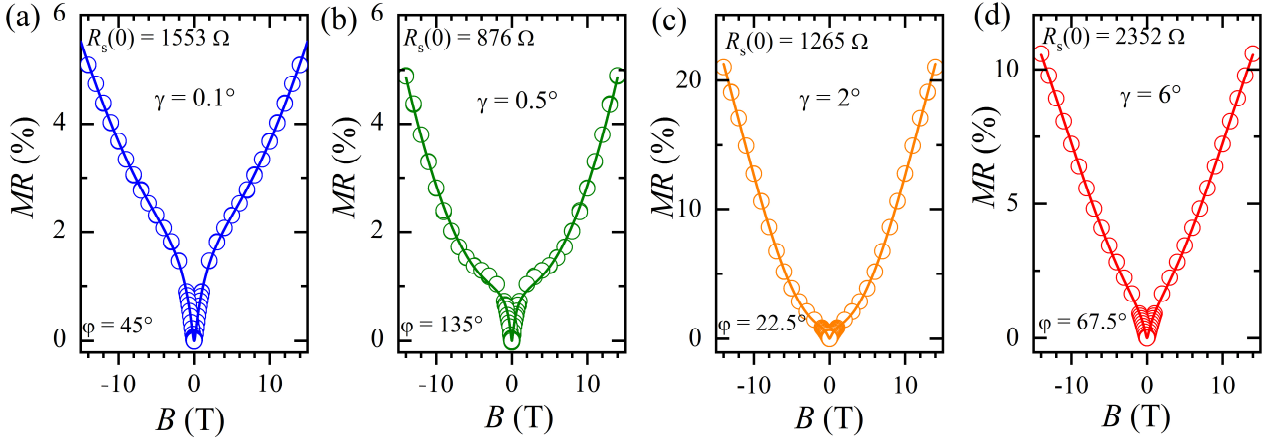

Figure 8: Magnetoresistance MR versus magnetic field B at T = 2 K of AO/STO heterostructures prepared on vicinal STO-substrates with miscut angle γ = 0.1° (a), 0.5° (b), 2° (c), and 6° (d). Solid lines are fits to the data points, see text. B was applied normal to the film surface. Measurements were carried out for current flow direction at φ close to φ*, see text. The zero-field sheet resistance $R_s(0)$ is indicated.

Fitting MR(B) with respect to WAL provides the characteristic field strengths for SOC, i. e., the inelastic field $B_i$ and the spin-orbit field $B_{so}$, which are related to the inelastic- and spin-orbit electron relaxation times, respectively. In the upper part of Fig. 9 we show $B_i$ and $B_{so}$ versus the current flow direction φ for the heterostructures with γ = 0.1°, 0.5°, 2°, and 6°. Generally, $B_{so}$ is about one order of magnitude larger compared to $B_i$ and amounts to 0.5 – 2 T. In addition, in contrast to $B_i$, the spin-orbit field displays φ-dependent and anisotropic behavior with a maximum value at φ ≈ φ*, the angle where $R_s(\varphi) = R_{max}$, see Fig. 5. The peaking of $B_{so}$ at φ*, indicated by arrows in Fig. 9, hints to an additional contribution to the spin-orbit relaxation time. Rashba-type SOC is well described by the D`yakonov-Perel (DP) spin-relaxation mechanism [49]. However, enhanced electron scattering by impurities may also lead to spin-flipping and hence spin-relaxation, known as the Elliott-Yafet (EY) spin-relaxation mechanism [50,51]. These are common spin-relaxation mechanisms in interface induced Rashba type spin-orbit coupling systems. The DP mechanism arises in systems that lack inversion symmetry whereas the EY mechanism originates from spin-orbit coupling induced spin dephasing due to electron-phonon coupling or interfacial defects [27]. Both type, DP and EY, can be identified by the relation between spin scattering timescale $\tau_{so}$ which is proportional to $1/B_{so}$ and momentum scattering timescale $\tau_p$ which is proportional to the electron mobility μ. For dominant DP mechanism, $\tau_{so}$ is proportional to $1/\tau_p$ and hence $B_{so}$ scales with μ whereas for dominant EY, $\tau_{so}$ linearly depends on $\tau_p$ which results in $B_{so}$ proportional to 1/μ.

In the lower part of Fig. 9 we have plotted the inverse of the charge carrier mobility 1/μ versus φ as deduced from $R_s(\varphi)$. Although the only few data points do not allow for a satisfactory scaling analysis of $\Delta B_{so}$ versus 1/μ, the maximum of 1/μ(φ) clearly coincidences with the maximum of $B_{so}$ at φ ≈ φ* which indicates distinct correlation of $\Delta B_{so}$ with 1/μ. Former measurements on AO/STO heterostructures grown on standard STO substrates displayed similar behavior [18]. Therefore, the peaking of $B_{so}$ is very likely related to enhanced electron scattering by impurities such as step-edges or dislocation lines of the vicinal STO substrate. In that context, the minima of $B_{so}(\varphi)$, indicated by the dashed lines in Fig. 9, reflect the spin-orbit field that is dominated by DP spin relaxation. Indeed, the minima of $B_{so}$ decreases with increasing γ for γ < 6°, whereas the minima of 1/μ (see dashed lines in the lower panel of Fig. 9) increase. That behavior suggests, that the minimum $B_{so}$ values are correlated to µ and not to 1/µ, as expected for DP spin relaxation. So, the spin-orbit field $B_{so}$ seems to be dominated by DP spin relaxation but also comprises contribution from defect-induced EY spin relaxation.

For $\gamma = 6°$, $B_{so}$ is quite large with respect to the rather low µ and seems to display a deviation from the behavior described above. This is also the case for the inelastic field. $B_i(\varphi)$ is almost constant and displays only noticeable $\varphi$-dependence for the sample with the largest miscut angle $\gamma = 6°$. Here, the $\varphi$-dependence of $B_i(\varphi)$ appears to be similar to that of $B_{so}(\varphi)$, indicating significant changes in the inelastic scattering process. The changed behavior of $B_{so}$ and $B_i$ for $\gamma = 6°$ are likely further indications for the enhanced electronic anisotropy. One should be aware that the functional field-dependence of the characteristic field strengths may change in the crossover regime from 2D to quasi 1D electronic transport, making a quantitative discussion of $B_i$ and $B_{so}$ critical.

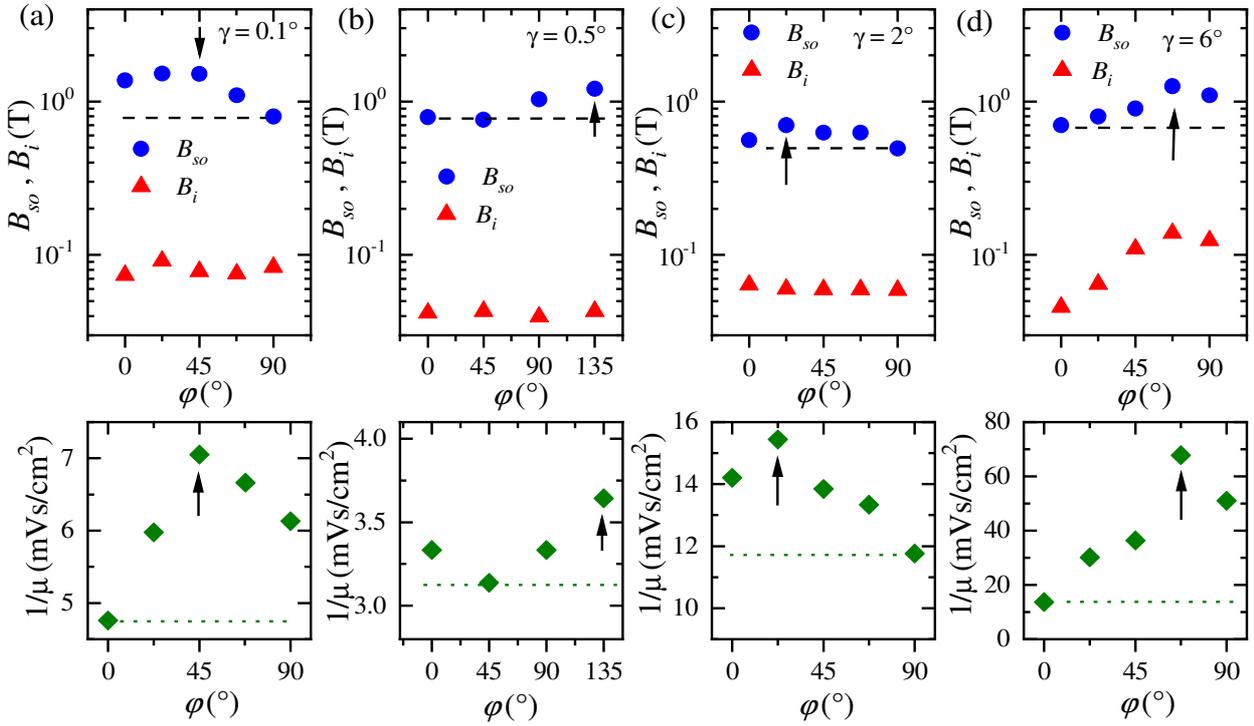

Figure 9: Inelastic field $B_i$ and spin-orbit field $B_{so}$ (upper part) and the inverse of the charge carrier mobility 1/µ (lower part) versus current flow direction $\varphi$ for AO/STO heterostructures on vicinal STO-substrates with miscut angle $\gamma = 0.1°$ (a), 0.5° (b), 2° (c), and 6° (d). $B_{so}$ peaks close to the position of $\varphi^*$, where $R_s(\varphi)$ and hence 1/µ display maximum values, see arrows. The minimum values of $B_{so}$ and 1/µ are indicated by dashed and dotted lines, respectively.

From the results above we conclude, that the Rashba-type SOC, which is characterized by DP spin-relaxation mechanism, does not show significant anisotropic behavior with respect to the current flow direction $\varphi$ or the anisotropic electronic transport in AO/STO heterostructures. A distinct relation of the Rashba-type SOC to the miscut angle $\gamma$ of the vicinal STO substrates is not evident. Even though there is anisotropic behavior of $B_{so}$, the reason for that is primarily related to spin-flipping by impurity scattering, *i. e.*, EY spin-relaxation. The simultaneous occurrence of DP and EF spin-relaxation mechanisms hinder a disentanglement and a more accurate determination of $B_{so}$ with respect to Rashba-type specific DP spin-relaxation. Nevertheless, for $\gamma > 2°$, SOC displays deviations from standard behavior. Generally, the values of $B_{so}$ for the AO/STO heterostructures vary between 0.5 and 2 T and are well comparable to those observed for LAO-STO. Assuming similar SOC in both kinds of heterostructures, the polar mismatch at the interface, responsible for the build-in electric field and Rashba effect, is expected to be comparable alike. That may be the reason for similar sheet carrier concentration in AO-STO and LAO-STO.

**IV. Summary**

A systematic study on the influence of the vicinal substrate miscut on the anisotropic 2D electronic transport in AO/STO heterostructures was carried out. To this end, the vicinal miscut angle $\gamma$ of the Ti-STO substrates, which were used for the AO film deposition, were varied from 0.1° to 6°. Measurements of the in-plane sheet resistance $R_s(\varphi)$ of the 2D electron system were done for different current flow directions $\varphi$ using patterned microbridges. Anisotropic transport evolves below 30 K. Analysis of $R_s(\varphi)$ reveals impurity scattering by lattice dislocations of bulk STO and interfacial scattering by the step-edges due to the vicinal substrate miscut as the main reason for the anisotropic behavior. The anisotropic contribution to the resistance caused by interfacial scattering, $R_t$, systematically increases with increasing $\gamma$ and decreasing terrace width $w$. In comparison to a standard substrate miscut of 0.1° ($w \approx 224$ nm), the amplitude of $R_t$ is increased by about one order of magnitude for $\gamma = 6°$ ($w \approx 18$ nm). However, the total anisotropy $R_{max}/R_{min}$ only increases by a factor of 2.6. The influence of $\gamma$ on $R_{max}/R_{min}$ is notably reduced by the occurrence of step-bunching and lattice-dislocations in the STO substrate material. Step-bunching limits the terrace width, and anisotropic scattering by lattice dislocation may diminish or even overcompensate the influence of interfacial scattering. Therefore, an accurate tuning of the anisotropic transport by $\gamma$ is hampered.

The magnetoresistance $MR$ increases with decreasing temperature $T$ and increasing magnetic field $B$ and reaches values up to about 25%. The positive $MR$ is well described by classical Lorentz scattering and weak-antilocalization (WAL) of the correlated 2D electron system. In addition, $MR$ becomes anisotropic at large fields ($B > 3$ T). With increasing $\gamma$ the anisotropy increases and the field-dependence of $MR$ becomes more linear, indicating enhanced disorder and anisotropic behavior of the 2D electron system. From the WAL, the inelastic field $B_i$ and the spin-orbit field $B_{so}$ which are related to the corresponding electron relaxation times were deduced. The Rashba-type SOC, which is characterized by the D`yakonov-Perel (DP) spin-relaxation mechanism, does not show significant anisotropic behavior with respect to the current flow direction $\varphi$ or the anisotropic electronic transport in AO/STO. A distinct relation of the Rashba-type SOC to the miscut angle $\gamma$ of the vicinal STO substrates is thus not evident. Even though there is an anisotropic behavior of $B_{so}$, the reason for that is primarily related to spin-flipping by impurity scattering, *i. e.*, Elliott-Yafet (EY) spin-relaxation mechanism. The charge carrier mobility $\mu$ and $R_s$ affect $B_{so}$ which prevents a more detailed analysis of the Rashba-type SOC. For $\gamma = 6°$, SOC displays deviations from standard behavior which might be taken as an indication for a beginning crossover from a 2D to a highly anisotropic (quasi 1D) electronic transport. The values of $B_{so}$ for the AO/STO heterostructures vary between 0.5 and 2 T and are well comparable to those observed for LAO/STO. Assuming similar SOC coupling in both kinds of heterostructures, the polar mismatch at the interface, responsible for the build-in electric field and Rashba effect, is expected to be comparable alike.

**Acknowledgements:**

DF and RS acknowledge the Deutsche Forschungsgemeinschaft (DFG) for financial support, Grant No. FU 457/2-1. The authors are grateful to R. Thelen and the Karlsruhe Nano Micro Facility (KNMF) for technical support. We also acknowledge D. Gerthsen and M. Meffert from the laboratory for electron microscopy (LEM) for transmission electron microscopy analysis of our samples. Thanks to R. Eder and J. Schmalian for fruitful discussion.